\documentclass[aps,prb]{revtex4}
\usepackage{amsmath}
\usepackage{graphicx}
\usepackage{epstopdf}
\usepackage{bm}
\begin{document}

\title{Spin waves in a skyrmion crystal}
\author{Olga Petrova}
\author{Oleg Tchernyshyov}
\affiliation{Department of Physics and Astronomy, Johns Hopkins University, Baltimore, Maryland 21218, USA}

\begin{abstract}
We derive the spectrum of low-frequency spin waves in skyrmion crystals observed recently in noncentrosymmetric ferromagnets. We treat the skyrmion crystal as a superposition of three helices whose wavevectors form an equilateral triangle. The low-frequency spin waves are Goldstone modes associated with displacements of skyrmions. Their dispersion is determined by the elastic properties of the skyrmion crystal and by the kinetic terms of the effective Lagrangian, which include both kinetic energy and a Berry-phase term reflecting a nontrivial topology of magnetization. The Berry-phase term acts like an effective magnetic field, mixing longitudinal and transverse vibrations into a gapped cyclotron mode and a twist wave with a quadratic dispersion. 
\end{abstract}

\maketitle

\section{Introduction}
\label{sec:intro}

Baby skyrmions are magnetic textures conjectured to exist in two-dimensional Heisenberg ferromagnets. \cite{JETPLett.22.503} Like domain walls and vortices, skyrmions are stable for topological reasons.  A localized magnetic texture, parametrized by the unit vector field $\hat{\mathbf m}(\mathbf r)$, has a quantized topological charge 
\begin{equation}
n = 
\int d^2r \, \frac{\hat{\mathbf m} \cdot (\partial_x \hat{\mathbf m} \times \partial_y \hat{\mathbf m})}{4\pi}.
\end{equation}
A state with a single skyrmion ($n = \pm 1$) cannot be continuously deformed into a uniform ground state ($n = 0$). While skyrmions are metastable excitations in the pure Heisenberg model, the presence of additional interactions may lower their energy cost and create a ground state with a finite concentration of skyrmions.  Bogdanov and collaborators pointed out that a Lifshitz invariant $\hat{\mathbf m} \cdot (\nabla \times \hat{\mathbf m})$ in the free energy may stabilize skyrmion-like line defects in a three-dimensional ferromagnet with easy-axis anisotropy. \cite{JETP.68.101, JETPLett.62.247, Nature.442.797} Such a term is allowed in a crystal without the inversion symmetry.  

\begin{figure}
\includegraphics[width=0.4\columnwidth]{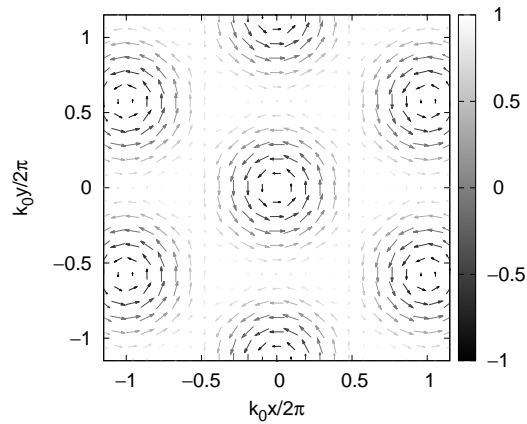}
\caption{Distribution of magnetization $\mathbf M(\mathbf r)$ of a skyrmion crystal in a plane $z = \mathrm{const}$ with magnetic field applied normal to that plane. The crystal is a superposition of three helices with wavenumbers given by Eq.~(\ref{eq:k0-abc0}). The vector plot shows components $M_x$ and $M_y$; shades of gray encode $M_z$ (arbitrary units).}
\label{fig:3-helices}
\end{figure}

Recent experiments\cite{Science.323.915, PhysRevB.81.041203, Nature.465.901, NatMater.10.106} have revealed ordered arrays of skyrmion lines in three-dimensional MnSi and Fe$_{1-x}$Co$_x$Si and arrays of skyrmions in thin films of Fe$_{0.5}$Co$_{0.5}$Si and FeGe. A common theme of these materials is the lack of the inversion symmetry.  Thin films are a particularly favorable environment for ``skyrmion crystals'' (SkX).  The stabilizing factor is likely dipolar interactions, which tend to favor inhomogeneous states such as the closely related bubble phase in a two-dimensional Ising ferromagnet. \cite{PhysRevB.26.325} 

The dynamics of magnetization in the SkX phase is an interesting question in its own right. Zang \textit{et al.} approached it by treating skyrmions as point-like particles. \cite{PhysRevLett.107.136804} Because of a nontrivial topology of a skyrmion, its kinematics are similar to that of a charged particle in a uniform magnetic field.  Moving skyrmions are subject to a Lorentz-like force that is proportional to the skyrmion velocity and is thus more important than inertia in the limit of slow motion. (A similar effect is well known for magnetic vortices. \cite{Science.304.420}) Like a Wigner crystal in magnetic field, \cite{JETP.51.148, JPSJ.41.3431} an SkX exhibits a quadratic phonon dispersion, $\omega \sim q^2/2m$. 

In the SkX phase, skyrmions form a dense lattice with the distance between skyrmions comparable to their size. \cite{Nature.465.901, NatMater.10.106} In this case, treating skyrmions as point particles may not be justified. From a complementary perspective, the SkX phase is a spin-density wave (SDW) depicted in Fig.~\ref{fig:3-helices}. It is a superposition of uniform magnetization parallel to the applied field $\mathbf H$ and three phase-locked helices whose wavevectors, 
\begin{equation}
\mathbf k_{0a} = k_0(1,0,0),\quad
\mathbf k_{0b} = k_0(-1/2,\sqrt{3}/2,0),\quad
\mathbf k_{0c} = k_0(-1/2,-\sqrt{3}/2,0).
\label{eq:k0-abc0}
\end{equation}
form an equilateral triangle in the plane normal to $\mathbf H$.\cite{Science.323.915} In the vicinity of the Curie point, higher harmonics are suppressed because they are not soft modes. This makes the SDW picture a more accurate starting point. In light of that, we set out to characterize low-frequency excitations of the SkX phase in the SDW limit.  The calculation of the spin-wave spectrum is nontrivial even for a single helix. \cite{PhysRevB.73.054431, PhysRevB.73.174402} To make the paper self-contained, we present a simple derivation of the spin-wave spectrum in a single helix for a three-dimensional ferromagnet in Sec.~\ref{sec:helix}. We then discuss the case of three coupled helices in Sec.~\ref{sec:skx}. 

The results can be briefly summarized as follows. The SkX (Fig.~\ref{fig:3-helices}) breaks the symmetry of translations in the $xy$ plane; translations in the $z$ direction remain a good symmetry. Thus low-energy excitations of the crystal are associated with displacements in the $xy$ plane, $\mathbf u(\mathbf r) = (u_x, u_y, 0)$. The potential energy of these displacements is expected to be the same as that of an isotropic two-dimensional (columnar) solid, 
\begin{equation}
\mathcal U =  
	\sum_{i=x,y}\sum_{j=x,y} 
	\left(
		\frac{\lambda}{2} u_{ii} u_{jj} + \mu u_{ij} u_{ij}
	\right)
	+ \frac{B (\partial_z^2 \mathbf u)^2}{2}.
\label{eq:u-solid}
\end{equation}
Here $u_{ij} = (\partial_i u_j + \partial _j u_i)/2$ are components of the strain tensor in the $xy$ plane, and $\lambda$ and $\mu$ are the elastic Lam{\'e} coefficients. \cite{Chaikin} Displacements inhomogeneous along the $z$-axis bend the crystal, hence an elastic energy proportional to $(\partial_z^2 \mathbf u)^2$ with a bending modulus $B$. The dispersion of elastic waves depends on the kinetic terms in the Lagrangian. By analogy with ordinary solids, one might expect a kinetic-energy term $\rho (\dot{\mathbf u})^2/2$, which is indeed present. Excitations in a columnar solid with these properties would be longitudinal and transverse sound waves with a linear dispersion for waves propagating in the plane of the crystal and a quadratic one for waves propagating normal to the plane. However, the nontrivial topology of magnetization in a unit cell adds a Berry-phase term $\mathcal{B} \dot{u}_x u_y$ to the Lagrangian of the SkX. Because this term is linear in the velocity, it dominates over the kinetic energy in the limit of slow motion. The resulting Lagrangian is
\begin{equation}
\mathcal L = \mathcal{B} \dot{u}_x u_y 
	+ \frac{\rho (\dot{\mathbf u})^2}{2} 
	- \sum_{i=x,y}\sum_{j=x,y} 
	\left(
		\frac{\lambda}{2} u_{ii} u_{jj} + \mu u_{ij} u_{ij}
	\right)
	- \frac{B (\partial_z^2 \mathbf u)^2}{2},
\label{eq:main-result-intro}
\end{equation}
At the lowest frequencies, transverse and longitudinal spin waves are mixed and exhibit a quadratic dispersion, $\omega \sim q^2/2m$, in agreement with the result obtained in the limit of point-like skyrmions. \cite{PhysRevLett.107.136804} 

With the sole exception of the Berry-phase term, the coupling constants of the SkX are set by the physics of a single helix. It is interesting that the anharmonic coupling that locks the phases of the helices does not enter the Lagrangian (\ref{eq:main-result-intro}). This coupling constant sets an upper limit of frequencies for which the three locked helices move together as a solid. 

\section{Single helix}
\label{sec:helix} 

\subsection{Static solutions}
\label{sec:helix-static}

We consider an isotropic ferromagnet in three dimensions with the magnetization field $\mathbf M(\mathbf r)$. The Landau free-energy functional is 
\begin{equation}
F[\mathbf M(\mathbf r)] = \int d^3r \left[A (\nabla \mathbf M)^2 + a M^2 + c M^4 
+ D \mathbf M \cdot (\nabla \times \mathbf M)\right],
\label{eq:F-of-M}
\end{equation}
where $A$ is the exchange coupling and $(\nabla \mathbf M)^2 \equiv \sum_i \sum_j (\partial_i M_j)^2$. The Dzyaloshinskii-Moriya coupling $D \mathbf M \cdot (\nabla \times \mathbf M)$ breaks the inversion symmetry and is only allowed in magnets with a noncentrosymmetric lattice. 

\begin{figure}
\includegraphics[width=0.4\columnwidth]{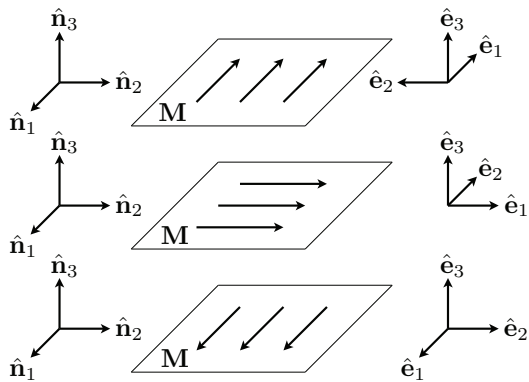}
\caption{The global frame $\{\hat{\mathbf n}_1, \hat{\mathbf n}_2, \hat{\mathbf n}_3\}$ and the local frame $\{\hat{\mathbf e}_1, \hat{\mathbf e}_2, \hat{\mathbf e}_3\}$ of a single helix, Eq.~(\ref{eq:helix}).}
\label{fig:frames}
\end{figure}

The quadratic part of the free energy is diagonalized by Fourier transform,
\begin{equation}
F[\mathbf M(\mathbf r)] 
= \sum_{\mathbf k} M^{\alpha*}_{\mathbf k} \Lambda^{\alpha\beta}_{\mathbf k} M^\beta_\mathbf k, 
\end{equation}
with a coupling matrix
\begin{equation}
\Lambda^{\alpha\beta}_{\mathbf k} = (a + Ak^2) \delta^{\alpha\beta} - iD \varepsilon^{\alpha\beta\gamma} k^\gamma.
\end{equation}
Its lowest eigenvalue, $a + Ak^2 - Dk$, is minimized by wavenumber $k_0 = D/2A$. When $a$ drops below the critical value $a_c = D^2/4A$, the ground state is a helical SDW, 
\begin{equation}
\mathbf M (\mathbf r) = M_0 
	[ \hat {\mathbf n}_1 \cos{(\mathbf k_0 \cdot \mathbf r)}
	+ \hat {\mathbf n}_2 \sin{(\mathbf k_0 \cdot \mathbf r)}],
\label{eq:helix}
\end{equation}
where $\mathbf k_0 = k_0 \hat {\mathbf n}_3$ and $M_0^2 = (a_c-a)/(2c)$. The unit vectors $\{\hat {\mathbf n}_1, \hat {\mathbf n}_2, \hat {\mathbf n}_3\}$ are mutually orthogonal and form a right-handed triplet.  We will also find it convenient to use a local frame set by the unit vectors $\{\hat {\mathbf e}_1, \hat {\mathbf e}_2, \hat {\mathbf e}_3\}$, where
\begin{equation}
\hat {\mathbf e}_1 = \hat {\mathbf n}_1 \cos{(\mathbf k_0 \cdot \mathbf r)}
	+ \hat {\mathbf n}_2 \sin{(\mathbf k_0 \cdot \mathbf r)},
\quad
\hat {\mathbf e}_2 = -\hat {\mathbf n}_1 \sin{(\mathbf k_0 \cdot \mathbf r)}
	+ \hat {\mathbf n}_2 \cos{(\mathbf k_0 \cdot \mathbf r)},
\quad 
\hat {\mathbf e}_3 = \hat {\mathbf n}_3.
\label{eq:local-frame}
\end{equation}
Here $\hat {\mathbf e}_1$ is parallel to the local direction of magnetization in equilibrium, the pair $(\hat {\mathbf e}_1, \hat {\mathbf e}_2)$ defines the plane of the helix, and $\hat {\mathbf e}_3$ is parallel to its wavevector $\mathbf k_0$. The orientations of the local and global frames are shown in Fig.~\ref{fig:frames}.

\subsection{Spin waves}

Small deviations from equilibrium can be described by two angles parametrizing twists of magnetization in the plane of the helix ($\alpha$) and out of the plane ($\beta$):
\begin{equation}
\mathbf M/M_0 = \hat {\mathbf e}_1 \left[1-(\alpha^2+\beta^2)/2\right]
 	+ \alpha \hat {\mathbf e}_2 + \beta \hat {\mathbf e}_3.
\label{eq:sh-gen}
\end{equation}
For convenience, we align the coordinate axes $x$, $y$, and $z$ with the vectors $\hat {\mathbf n}_1$, $\hat {\mathbf n}_2$, and $\hat {\mathbf n}_3$. To the second order in the twist angles, the energy cost is
\begin{subequations}
\label{eq:U-ugly}
\begin{eqnarray}
\mathcal U &=& DM_0^2 (-\partial_x \sin{k_0 z} + \partial_y \cos{k_0 z})\beta 
	\label{eq:U-ugly-a}\\
 && + A M_0^2 \left[(\nabla \alpha)^2 + (\nabla \beta)^2 + k_0^2 \beta^2 \right]
 	\label{eq:U-ugly-b}\\
 && + D M_0^2 [\beta (\partial_x \cos{k_0 z} + \partial_y \sin{k_0 z}) \alpha
    - \alpha (\partial_x \cos{k_0 z} + \partial_y \sin{k_0 z}) \beta].
    \label{eq:U-ugly-c}
\end{eqnarray}
\end{subequations}
The linear term (\ref{eq:U-ugly-a}) is a total divergence that reduces to a boundary term upon integration over the volume. It does not influence the physics in the bulk and can be ignored. 

The dynamics of spin waves can be obtained by constructing a Lagrangian for the fields $\alpha$ and $\beta$. The kinetic term comes from the Berry phase for a spin, $S_B = S\int (\cos{\theta}-1) \dot \phi \, dt$. \cite{Altland-Simon} Measuring the inclination angle $\theta$ from the local magnetization axis $\hat {\mathbf e}_1$ yields $\alpha = \theta \cos{\phi}$ and $\beta = \theta\sin{\phi}$ for infinitesimal $\alpha$ and $\beta$, so that the Berry-phase term in the Lagrangian becomes $\mathcal L_B = \mathcal J(\dot \alpha \beta - \alpha \dot \beta)/2$, or equivalently $\mathcal J \dot \alpha \beta$, where $\mathcal J = M_0/\gamma$ is the density of angular momentum and $\gamma$ is the gyromagnetic ratio. The resulting Lagrangian is 
\begin{equation}
L = L_B - \mathcal U 
	= \mathcal J \dot \alpha \beta 
	- A M_0^2 \left[(\nabla \alpha)^2 + (\nabla \beta)^2 + k_0^2 \beta^2 \right]
	- 2D M_0^2 \beta (\partial_x \cos{k_0 z} + \partial_y \sin{k_0 z}) \alpha.
\label{eq:L}
\end{equation}
The Euler-Lagrange equations of motion are
\begin{subequations}
\begin{eqnarray}
-[\mathcal J \partial_t + 2D M_0^2 (\partial_x \cos{k_0 z} + \partial_y \sin{k_0 z})] \beta 
	&=& 2A M_0^2 \nabla^2 \alpha,
\label{eq:eom-a}\\
{}
[\mathcal J \partial_t + 2D M_0^2 (\partial_x \cos{k_0 z} + \partial_y \sin{k_0 z})] \alpha 
	&=& 2A M_0^2 (\nabla^2 \beta - k_0 ^2 \beta).
\label{eq:eom-b}
\end{eqnarray}
\end{subequations}
The equations, though linear, are complicated by the presence of oscillatory terms reflecting  the breaking of the translational symmetry by the helix. Simple closed-form solutions can only be obtained for waves propagating along the direction of the helix $\hat {\mathbf n}_3$, $\alpha = \bar \alpha \cos{(\omega t - q_{||} z)}$, $\beta = \bar \beta \sin{(\omega t - q_{||} z)}$, with the dispersion 
\begin{equation}
\omega = (2A M_0^2/\mathcal J) \sqrt{k_0^2 q_{||}^2 + q_{||}^4}.
\label{eq:disp-longitudinal}
\end{equation}
For short wavelengths, $q_{||} \gg k_0$, we recover the quadratic dispersion of magnons in a Heisenberg ferromagnet. For wavelengths longer than the pitch of the helix, the dispersion becomes linear, $\omega \sim s q_{||}$, with the wave velocity $s = 2 A M_0^2 k_0/\mathcal J = \gamma M_0 D$. It is worth noting that the wave velocity is independent of the exchange coupling $A$.

\subsection{Long-wavelength approximation}

Further progress can be made by analyzing spin waves in the long-wavelength limit. The helical modulation mixes waves with wavevectors $\mathbf q$ and $\mathbf q \pm \mathbf k_0$, but the effects of the mixing can be controlled in the limit $q \ll k_0$. 

As usual, we expect that low-frequency waves are associated with the generators of broken symmetries, in this case the global rotational symmetry SO(3) and translations along the helix axis $\hat {\mathbf n}_3$. Zero modes, which leave the energy of the system invariant, can also be obtained by requiring that the first variation of energy vanish: $\delta \int \mathcal U \, dV = 0$. The general solution is an infinitesimal global rotation parametrized by three angles $\{\phi_i\}$ about the global axes $\{\hat {\mathbf n}_i\}$:
\begin{equation}
\alpha = -\phi_1 k_0 y + \phi_2 k_0 x + \phi_3,
\quad
\beta = \phi_1 \sin{k_0 z} - \phi_2 \cos{k_0 z}.
\label{eq:0-modes}
\end{equation}
Note that $\phi_3$ can also be viewed as a translation of the helix through distance $-\phi_3/k_0$ along its axis $\hat {\mathbf n}_3$. 

Although there are three parameters describing an infinitesimal global rotation, only one of them---$\phi_3$---is associated with a Goldstone mode. Radzihovsky and Lubensky \cite{PhysRevE.83.051701} pointed out that the other two modes become ``massive'' via a Higgs-like mechanism similar to that in smectics A. \cite{SSC.10.753}

The explicit form of the zero modes (\ref{eq:0-modes}) shows that soft modes of the helix are situated near wavevectors 0 and $\mathbf k_0$.  We therefore focus on the respective Fourier components of $\alpha$ and $\beta$: 
\begin{equation}
\alpha(\mathbf r) = \alpha_0 + \alpha_1 \cos{k_0 z} 
	+ \alpha_2 \sin{k_0 z},
\quad
\beta(\mathbf r) = \beta_0 + \beta_1 \cos{k_0 z} 
	+ \beta_2 \sin{k_0 z},
\end{equation}
where fields $\alpha_i(\mathbf r)$ and $\beta_i(\mathbf r)$ vary slowly in space. We express the Lagrangian (\ref{eq:L}) in terms of these. After averaging over spatial regions large compared to the helix period, oscillatory terms such as $\cos{2 k_0 z}$ vanish and we obtain
\begin{subequations}
\begin{eqnarray}
\mathcal L &=& \dot \alpha_0 \beta_0 
	- (\nabla \alpha_0 + \bm \beta)^2 
	- \beta_0^2 
	- \frac{(\nabla \bm \beta)^2}{2} 
\label{eq:L-cute-a}
\\
	&&+ \frac{\dot {\bm \alpha} \cdot {\bm \beta}}{2} 
	- (\nabla \beta_0)^2 
	- \frac{\bm \alpha^2}{2} - \frac{(\nabla \bm \alpha)^2}{2}	
	+ 2 \bm \alpha \cdot \nabla \beta_0
	+ \frac{\bm \alpha \cdot \nabla \times \bm \alpha}{2}
	+ \frac{\bm \beta \cdot \nabla \times \bm \beta}{2}.
\label{eq:L-cute-b}
\end{eqnarray}
\end{subequations}
Here we introduced vectors $\bm \alpha = (\alpha_1, \alpha_2, 0)$ and $\bm \beta = (\beta_1, \beta_2, 0)$ and switched to natural units,
\begin{equation}
\mbox{length: } 1/k_0, 
\quad
\mbox{time: } \mathcal J/A M_0^2 k_0^2,
\quad
\mbox{energy: } A M_0^2/k_0.
\end{equation}
The term $(\nabla \bm \beta)^2$ is understood as $(\nabla \beta_1)^2 + (\nabla \beta_2)^2$.

The low-frequency dynamics of the soft mode $\alpha_0$ is determined by the terms displayed in Eq.~(\ref{eq:L-cute-a}). The rest of the terms (\ref{eq:L-cute-b}) provide corrections containing higher powers of the gradients and can be safely dropped in the long-wavelength limit. 

\begin{figure}
\includegraphics[width=0.32\columnwidth]{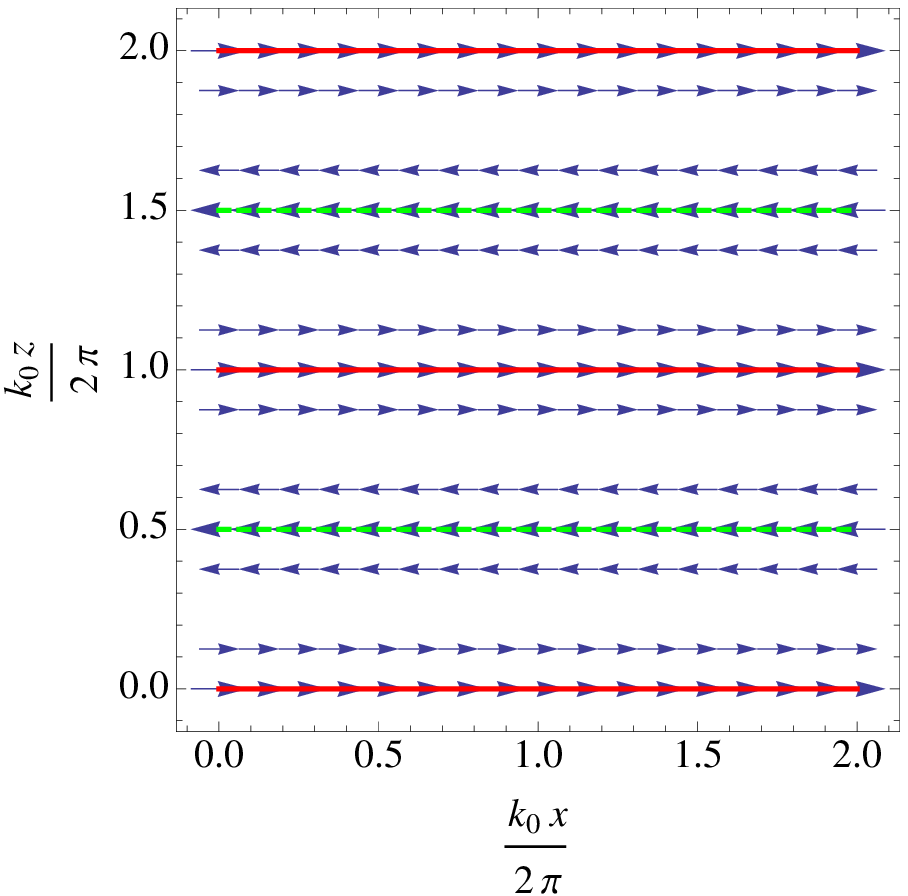}
\includegraphics[width=0.32\columnwidth]{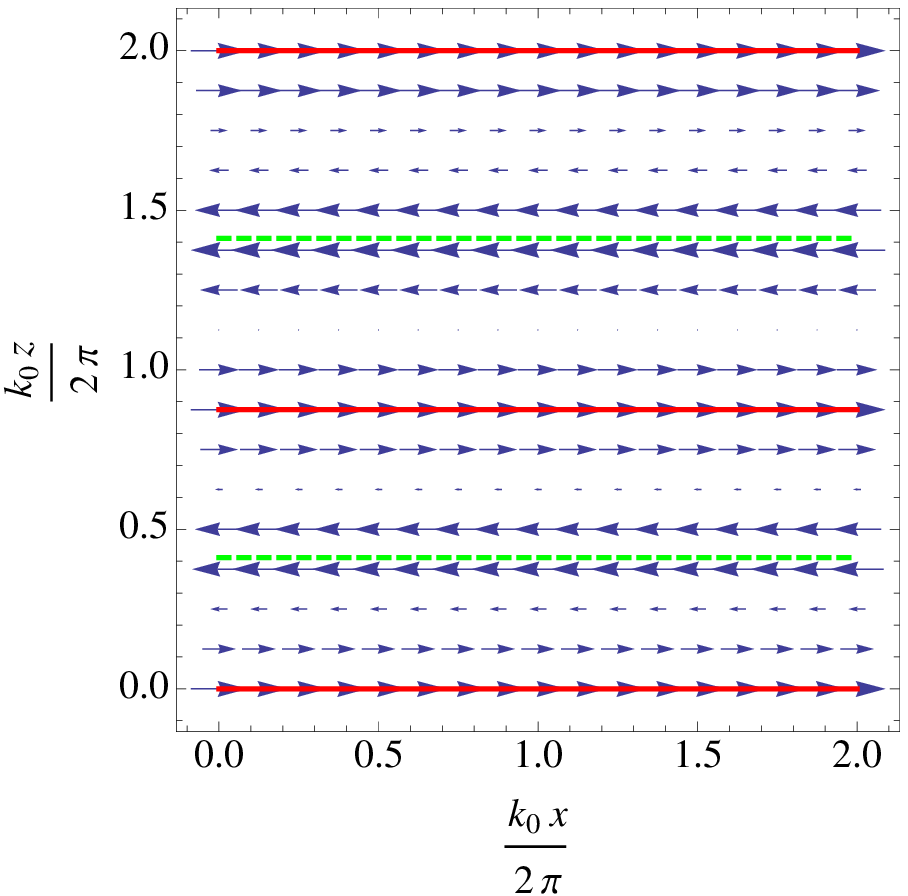}
\includegraphics[width=0.32\columnwidth]{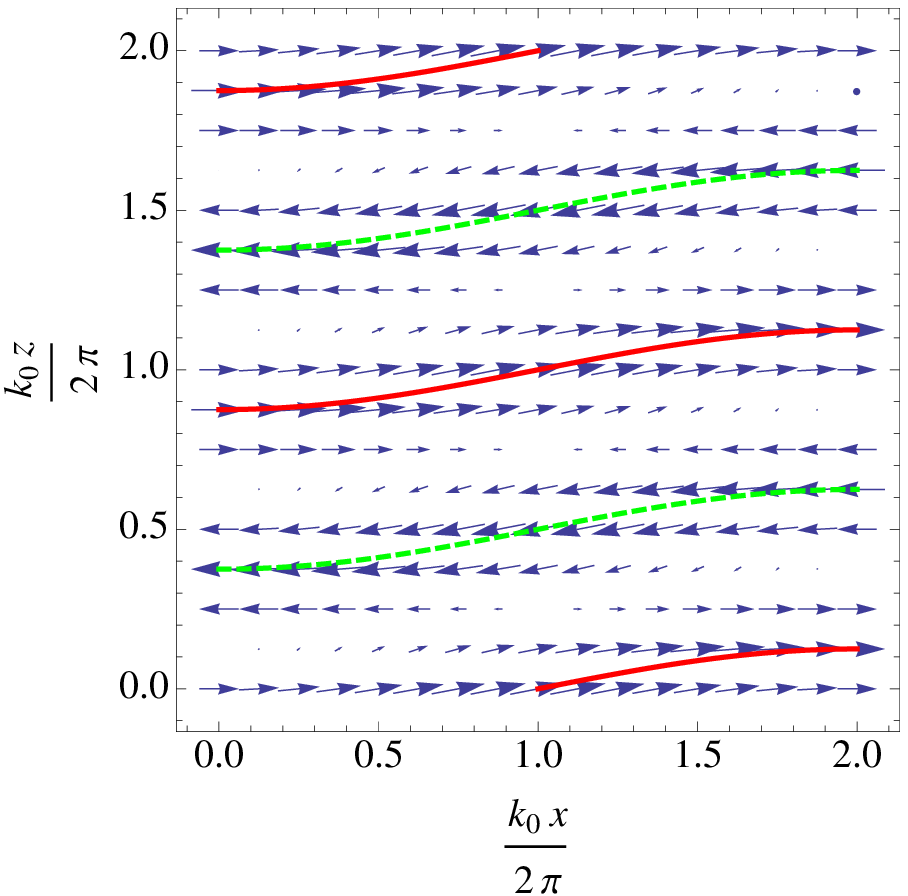}
\caption{Spin waves in a helical state. Left: a helix in equilibrium. The helix wavevector $\mathbf k_0$ is vertical. Solid red and dashed green lines are planes of constant phase $k_0 z = 2\pi n$ and $k_0 z = 2\pi (n+1/2)$. Center: a spin wave with the propagation vector $\mathbf q$ parallel to $\mathbf k_0$. Right: a spin wave with the propagation vector $\mathbf q$ perpendicular to $\mathbf k_0$. Planes of constant phase have a vertical displacement given by Eq.~(\ref{eq:u-helix}).}
\label{fig:helix-waves}
\end{figure}

Integrating out the hard field $\beta_0$ with the aid of its equation of motion, $\beta_0 = \dot \alpha_0/2$, generates a ``kinetic energy'' for the soft mode, $\dot \alpha_0^2/4$.  At low frequencies and long wavelengths, the ``gauge field'' $\bm \beta$ follows the gradient of the ``Higgs field'' $\alpha_0$ so that $\bm \beta \approx -\nabla_\perp \alpha_0$, where $\nabla_\perp = (\partial_x, \partial_y, 0)$ is the transverse part of the gradient. Integrating out $\bm \beta$ yields the following Lagrangian for the soft mode $\alpha_0$: 
\begin{equation}
\mathcal L = \frac{\dot \alpha_0^2}{4} - (\nabla_{||} \alpha_0)^2 
	- \frac{(\nabla_\perp^2 \alpha_0)^2}{2}. 
\label{eq:L-alpha0}
\end{equation}
Here $\nabla_{||} = \partial_z$ is the longitudinal part of the gradient.   

Low-frequency spin waves are slow twists of the magnetization in the helix plane described by the ``Higgs field'' $\alpha_0$, Fig.~\ref{fig:helix-waves}. Twists out of the plane, described by the ``gauge field'' $\bm \beta$, adjust to them as needed. Alternatively, the spin waves can be viewed as displacements of planes of constant phase of the helix along its propagation direction $\hat {\mathbf n}_3$,
\begin{equation}
u(\mathbf r) = -\alpha_0(\mathbf r)/k_0.
\label{eq:u-helix}
\end{equation}
Waves propagating along the helix direction are regular phonons, with the energy density $(\nabla_{||} \alpha_0)^2$. Waves propagating along the planes of constant phase are flexural modes, with a much lower energy density $(\nabla_\perp^2 \alpha_0)^2/2$. This results in a strongly anisotropic dispersion. With the units of length and time restored, we have
\begin{equation}
\omega = (2AM_0^2/\mathcal J)\sqrt{k_0^2 q_{||}^2 + q_{\perp}^4/2}.
\end{equation}
This expression agrees with the dispersion for the longitudinal direction, Eq.~(\ref{eq:disp-longitudinal}), in the long-wavelength limit.

\section{Skyrmion crystal}
\label{sec:skx}

\subsection{Three coupled helices}

The skyrmion-crystal phase is realized in the vicinity of the critical point in an applied magnetic field $\mathbf H$.\cite{Science.323.915}  The field induces a finite uniform magnetization $\overline{\mathbf M} = \chi \mathbf H$. In its presence, the quartic term in the free energy (\ref{eq:F-of-M}) generates an effective cubic term, 
\begin{equation}
\mathcal U_3 = 4 c \, (\overline{\mathbf M} \cdot \mathbf M) (\mathbf M \cdot \mathbf M),
\end{equation}
where the order parameter $\mathbf M$ contains only soft modes with wavenumbers near $k_0$ and excludes the uniform component of magnetization. By the standard argument,\cite{PhysRevLett.41.702, Chaikin, Science.323.915} the cubic term favors a magnetic state that is a superposition of three helices (\ref{eq:helix}) whose wavevectors $\mathbf k_{0a}$, $\mathbf k_{0b}$, and $\mathbf k_{0c}$ form an equilateral triangle orthogonal to the applied field $\mathbf H$. (It should be noted that the helix amplitude $M_0$ need not be the same as in the case of a single helix. Here we treat $M_0$ as a phenomenological parameter that determines the energetics of the soft modes.) The cubic term couples the phases of the three helices thereby generating an effective coupling between their soft modes $\alpha_{0a}$, $\alpha_{0b}$, and $\alpha_{0c}$:
\begin{equation}
\mathcal U_\mathrm{lock} = - \sigma \cos{(\alpha_{0a} + \alpha_{0b} + \alpha_{0c})}
\approx -\sigma + \frac{\sigma(\alpha_{0a} + \alpha_{0b} + \alpha_{0c})^2}{2}.
\end{equation}
where 
\begin{equation}
\sigma = \frac{8 c \overline{M} M_0^3}{A M_0^2 k_0^2} 
	= \frac{4 \overline{M}}{M_0 k_0^2 \xi_l^2}  
\label{eq:sigma}
\end{equation}
is a dimensionless phase-locking coupling and $\xi_l = \sqrt{A/(a_c-a)}$ is the longitudinal correlation length of magnetization. \cite{Kardar-fields}   

At the harmonic level, the three soft modes have the following Lagrangian:
\begin{equation}
\mathcal L = -\frac{\sigma(\alpha_{0a} + \alpha_{0b} + \alpha_{0c})^2}{2}
+ \sum_{i = a,b,c} 
	\left(
		\frac{\dot{\alpha}_{0i}^2}{4} 
		- (\nabla_{||i} \, a_{0i})^2 - \frac{(\nabla_{\perp i}^2 \, a_{0i})^2}{2}
	\right),
\end{equation}
where $\nabla_{||i} = \mathbf k_{0i} \cdot \nabla$ and $\nabla_{\perp i} = \mathbf k_{0i} \times \nabla$.  Thanks to the phase-locking coupling $\sigma$, one of the modes acquires a finite frequency $\omega_0 = \sqrt{6\sigma}$.  The two remaining Goldstone modes are translations $\mathbf u$ in the plane of the SkX.  (Translations normal to the plane do not alter the state of the system.) For convenience, we direct the $z$-axis along the applied field $\mathbf H$ and choose the $x$ and $y$-axes in such a way that 
\begin{equation}
\mathbf k_{0a} = (1,0,0),\quad
\mathbf k_{0b} = (-1/2,\sqrt{3}/2,0),\quad
\mathbf k_{0c} = (-1/2,-\sqrt{3}/2,0).
\label{eq:k0-abc}
\end{equation}
As in the case of a single helix, Eq.~(\ref{eq:u-helix}), a translation $\mathbf u$ yields a phase shift $\alpha_{0i} = - \mathbf k_{0i} \cdot \mathbf u$ for helix $i = a, b, c$. The resulting Lagrangian for the Goldstone modes $u_x$ and $u_y$ describes a columnar solid: 
\begin{equation}
\mathcal L =  
	\frac{\rho (\dot{\mathbf u})^2}{2} 
	- \sum_{i=x,y}\sum_{j=x,y} 
		\left(
			\frac{\lambda}{2} u_{ii} u_{jj} + \mu u_{ij} u_{ij}
		\right)
	- \frac{B (\partial_z^2 \mathbf u)^2}{2},
\label{eq:L-skx-solid}
\end{equation}
where $u_{ij} = (\partial_i u_j + \partial_j u_i)/2$ are components of strain in the $xy$ plane, $\rho = 3/4$ is the density, $B = 3/2$ is the bending modulus, and $\lambda = \mu = 3/4$ are the Lam{\'e} parameters. Note that the phase-locking coupling $\sigma$ does not enter the Lagrangian (\ref{eq:L-skx-solid}).  Instead, it sets the range of frequencies, 
\begin{equation}
\omega \lesssim \sqrt{6\sigma},
\label{eq:omega-range} 
\end{equation}
in which the texture behaves as a two-dimensional solid. At higher frequencies, the three helices decouple from one another.

Eigenmodes of the Lagrangian (\ref{eq:L-skx-solid}) come in two flavors: longitudinal waves, in which the displacement of the SkX $\mathbf u = (u_x, u_y, 0)$ is parallel to the in-plane component of the wavevector $\mathbf q_\mathrm{in} = (q_x,q_y,0)$, and transverse waves with $\mathbf u$ along $\hat{\mathbf z} \times \mathbf q_\mathrm{in} = (-q_y, q_x,0)$.  With the physical units restored, the dispersions are
\begin{equation}
\omega_l = (AM_0^2/\mathcal J)\sqrt{3k_0^2 q_\mathrm{in}^2 + 2q_\mathrm{out}^4},\quad
\omega_t = (AM_0^2/\mathcal J)\sqrt{k_0^2 q_\mathrm{in}^2 + 2q_\mathrm{out}^4},
\label{eq:disp}
\end{equation}
Here $q_\mathrm{out}=q_z$ is the out-of-plane component of the wavevector. This dispersion was obtained recently by Kirkpatrick, Belitz, and collaborators, \cite{PhysRevLett.104.256404, PhysRevB.82.134427} who treated the SkX as a two-dimensional solid (\ref{eq:L-skx-solid}). 

The first term in the effective Lagrangian (\ref{eq:L-skx-solid}) is proportional to the square of the velocity $\dot{\mathbf u}$ and thus can be viewed as kinetic energy of the SkX. As a result of that, waves propagating in the plane of the crystal have a linear dispersion $\omega = sq$ with the speed $s_l = AM_0^2 k_0 \sqrt{3}/\mathcal J = \gamma M_0 D \sqrt{3}/2$ for longitudinal waves and $s_t = s_l/\sqrt{3}$ for transverse ones.   

\subsection{Berry phase}

The low-frequency dispersions will be modified if the effective Lagrangian (\ref{eq:L-skx-solid}) acquires a Berry-phase term $\mathcal{B} \dot{u}_x u_y$. Being linear in the velocity, this term dominates over kinetic energy in the limit $\omega \to 0$. The effective Lagrangian then becomes
\begin{equation}
\mathcal L = \mathcal{B} \dot{u}_x u_y 
	+ \frac{\rho (\dot{\mathbf u})^2}{2} 
	- \sum_{i=x,y}\sum_{j=x,y} 
		\left(
			\frac{\lambda}{2} u_{ii} u_{jj} + \mu u_{ij} u_{ij}
		\right)
	- \frac{B (\partial_z^2 \mathbf u)^2}{2}.
\label{eq:main-result}
\end{equation}
We focus on waves propagating in the plane of the crystal, so that $\partial_z \mathbf u = 0$. The Berry-phase term acts like an effective magnetic field hybridizing longitudinal and transverse waves.  One of the resulting modes has a finite frequency,
\begin{equation} 
\omega_c = \mathcal{B}/\rho,
\label{eq:gyro-frequency}
\end{equation}
in the $q \to 0$ limit. It is the cyclotron mode that reflects the interplay of inertia and the Lorentz force acting on a charged particle in a magnetic field. The other mode has a quadratic dispersion, 
\begin{equation}
\omega = \omega_l \omega_t/\mathcal{B} = (q_\mathrm{in}^2/\mathcal{B})\sqrt{\mu(\lambda + 2\mu)}.
\label{eq:dispersion-quadratic}
\end{equation}
This result holds for sufficiently small in-plane wavenumbers, $q_\mathrm{in} \ll \mathcal{B}$. In the opposite limit, $q_\mathrm{in} \gg \mathcal{B}$, the motion becomes sufficiently fast for the Berry phase to be negligible relative to kinetic energy and the eigenmodes become longitudinal and transverse sound waves. 

The Berry-phase coupling $\mathcal{B}$ can be computed by moving the SkX along an infinitesimal contour of area $\Delta X \Delta Y$:
\begin{equation}
\mathbf u 
	\mapsto \mathbf u + (\Delta X, 0, 0) 
	\mapsto \mathbf u + (\Delta X, \Delta Y, 0)
	\mapsto \mathbf u + (0, \Delta Y, 0)
	\mapsto \mathbf u.
\end{equation}
In the process, the magnetization vector at a given point traces out a closed path on a sphere:
\begin{equation}
\mathbf M 
	\mapsto \mathbf M - \frac{\partial \mathbf M}{\partial x} \Delta X
	\mapsto \mathbf M - \frac{\partial \mathbf M}{\partial x} \Delta X 
		- \frac{\partial \mathbf M}{\partial y} \Delta Y 
	\mapsto \mathbf M - \frac{\partial \mathbf M}{\partial y} \Delta Y
	\mapsto \mathbf M.
\end{equation}
If magnetization is not confined to a single plane then the path on the sphere has a nonzero area and the spins acquire a Berry phase $S_B$ proportional to the area on the sphere and spin length. In this way we obtain 
\begin{equation}
\mathcal{B} = -\int_\Omega \frac{dx \, dy}{\Omega} \frac{M}{\gamma} \, 
	\hat{\mathbf m} \cdot 
		\left(
			\partial_x \hat{\mathbf m} \times \partial_y \hat{\mathbf m}
		\right),
\label{eq:g-def}
\end{equation}
where $\hat{\mathbf m} = \mathbf M/M$ is the unit vector parallel to magnetization and the integration is done over a unit cell of area $\Omega = 8 \pi^2 \sqrt{3}/k_0^2$. 

Deep in the ordered phase, the magnetization length $M$ is fixed and Eq.~(\ref{eq:g-def}) reduces to 
\begin{equation}
\mathcal{B} = - \frac{4\pi n M}{\gamma \Omega}, 
\label{eq:g}
\end{equation}
where 
\begin{equation}
n = \int_\Omega dx \, dy  \, 
	\frac{
		\hat{\mathbf m} \cdot
		\left(
			\partial_x \hat{\mathbf m} \times \partial_y \hat{\mathbf m}
		\right)
		}
		{4\pi}
\label{eq:n-def}
\end{equation}
is a topological charge of the unit cell known as the skyrmion number. In an SkX, $n = \pm 1$. In natural units, 
\begin{equation}
\mathcal{B} = \frac{nM}{2\pi M_0 \sqrt{3}}.
\label{eq:g-nat}
\end{equation}

Closer to the critical point, an SkX is a superposition of three phase-locked helices on top of uniform magnetization. In such a texture, magnetization length varies in space. \cite{arXiv:1001.1292}  Then $nM$ in Eqs.~(\ref{eq:g}) and (\ref{eq:g-nat}) should be understood as the average length of magnetization weighted with skyrmion density $\hat{\mathbf m} \cdot \left( \partial_x \hat{\mathbf m} \times \partial_y \hat{\mathbf m} \right)/(4\pi)$.

\section{Discussion}

We have derived a low-energy description of a skyrmion crystal viewed as a superposition of three phase-locked helices. \cite{Science.323.915} This approach is justified because skyrmion crystals are typically observed near the critical point, where soft components of magnetization have characteristic wavenumbers $k_0$. In this limit, the extent of a skyrmion is comparable to the interskyrmion separation, which means that these solitons cannot be treated as point particles. Viewing this magnetic texture as a superposition of three helical waves is a better starting point. 

Our main result is the Lagrangian for low-energy excitations of the skyrmion crystal, Eq.~(\ref{eq:main-result}). It is expressed in terms of the Goldstone modes of a columnar solid in three dimensions, deformations parametrized by a two-component displacement field $\mathbf u(\mathbf r) = (u_x, u_y, 0)$. The description is valid at sufficiently low frequencies, when the three helices behave as a cohesive solid. The upper limit is set by the strength of the anharmonic coupling constant (\ref{eq:sigma}) locking the phases of the helices. At intermediate frequencies, the dynamics is dominated by the inertia of the crystal, parametrized by the kinetic energy in Eq.~(\ref{eq:main-result}). In this regime, excitations are transverse and longitudinal vibrations with a linear dispersion, $\omega \sim s q$, for waves propagating in the plane of the crystal. 

At the lowest frequencies, the Berry-phase coupling mixes longitudinal and transverse waves. One of the mixed modes acquires a quadratic dispersion, $\omega \sim q^2/2m$ (\ref{eq:dispersion-quadratic}), as proposed by Zang \emph{et al.} \cite{PhysRevLett.107.136804} The other mode has a finite  frequency (\ref{eq:gyro-frequency}) in the $q \to 0$ limit. This mode corresponds to cyclotron motion of skyrmions: the Lagrangian (\ref{eq:main-result}) indicates that skyrmions behave as massive particles with density $\rho$ in an effective magnetic field of strength $\mathcal B$. The skyrmion mass density in standard units is $\rho = 3 \mathcal J^2/(4 A M_0^2)$. The mass of a skyrmion $m_s$ can be compared to the mass of a magnon $m_m$ in the same ferromagnetic material (without the Dzyaloshinskii-Moriya coupling): 
\begin{equation}
\frac{m_s}{m_m} \approx \frac{\ell}{a} \frac{1}{(k_0 a)^2},
\end{equation}
where $\ell$ is the length of the skyrmion line (in a three-dimensional crystal) and $a$ is the lattice spacing of the material. 

The observability of the skyrmion cyclotron mode depends on the strength of the skyrmion crystal measured by the phase-locking coupling $\sigma$ (\ref{eq:sigma}). The cyclotron should be in the range of frequencies (\ref{eq:omega-range}), where the texture behaves as a two-dimensional crystal, rather than three independent helices. If the crystal is too weak, $\sigma \lesssim \mathcal{B}^2/6$, the Lagrangian (\ref{eq:main-result}) only applies in the limit of slow motion, where inertia of the skyrmion crystal can be neglected. In that case, the density term may be entirely omitted.

It is interesting to compare the spectra of low-energy excitations of skyrmion crystals in ferromagnets and in systems exhibiting the quantum Hall effect (QHE). \cite{PhysRevB.47.16419} In QHE systems, the Coulomb interaction between charged skyrmions in two dimensions causes the  Lam{\'e} parameter $\lambda$ to diverge as $q^{-1}$ at low wavenumbers, which leads to the $q^{3/2}$ magnetophonon dispersion. \cite{PhysRevB.54.16838, PhysRevB.58.10634} The cyclotron mode (\ref{eq:gyro-frequency}) is absent because the mass density $\rho$ is small. \cite{PhysRevB.54.16838} This is similar to our case when a small $\rho$ pushes the cyclotron frequency outside the frequency range (\ref{eq:omega-range}) in which the texture can be thought of as a skyrmion crystal. In the absence of spin-orbital coupling, QHE skyrmion crystals get an additional Goldstone mode with a linear dispersion, associated with the spontaneous breaking of the global SO(2) symmetry of spin rotations. \cite{PhysRevLett.78.4825, PhysRevB.58.10634} In our system, such spin waves are coupled to translations of the skyrmions and are therefore gapped via the Higgs mechanism. \cite{PhysRevE.83.051701}

Although in this paper we have focused on a three-dimensional ferromagnet,\cite{Science.323.915, PhysRevB.81.041203} these results can be readily extended to thin films,\cite{Nature.465.901, NatMater.10.106} in which long-range dipolar interactions make the SkX phase particularly stable. 

\subsubsection*{Note added in proof}

A recent numerical study by Mochizuki\cite{arXiv:1111.5667} provides evidence for two branches of spin waves in a skyrmion crystal. 

\section*{Acknowledgments}

We thank D. Belitz, Z. H. Hao, C. Timm, and Y. Wan for useful comments. This work was supported in part by the US National Science Foundation under Award No. DMR-1104753.

\bibliographystyle{apsrev}
\bibliography{skyrmions,micromagnetics}

\end{document}